\documentclass[journal]{IEEEtran}
\usepackage{amssymb}
\usepackage{algorithm}
\usepackage{algorithmic}
\usepackage{array}
\usepackage{cite}
\usepackage{url}
\usepackage{caption}
\usepackage{float}
\usepackage{subfigure}
\usepackage{enumerate}
\usepackage{verbatim}
\usepackage{graphicx}
\usepackage{amsmath}
\usepackage{bbm}
\usepackage{booktabs}
\usepackage{epstopdf}
\usepackage{diagbox}
 \usepackage{color}

\title{Sneak Path Interference-Aware Adaptive Detection and Decoding for Resistive Memory Arrays}

\author{Panpan Li, Kui Cai, Zhen Mei, Guanghui Song}

\begin{document}

\maketitle

\begin{abstract}
Resistive random-access memory (ReRAM) is an emerging non-volatile memory technology for high-density and high-speed data storage. However, the sneak path interference (SPI) occurred in the ReRAM crossbar array seriously affects its data recovery performance. In this letter, we first propose a quantized channel model of ReRAM, based on which we design both the one-bit and multi-bit channel quantizers by maximizing the mutual information of the channel. A key channel parameter that affects the quantizer design is the sneak path occurrence probability (SPOP) of the memory cell. We first use the average SPOP calculated statistically to design the quantizer, which leads to the same channel detector for different memory arrays. {  We then adopt the SPOP estimated separately for each memory array for the quantizer design, which is generated by an effective channel estimator and through an iterative detection and decoding scheme for the ReRAM channel.} This results in an array-level SPI-aware adaptive detection and decoding approach. Moreover, since there is a strong correlation of the SPI that affects memory cells in the same rows/columns than that affecting cells in different rows/columns, we further derive a column-level scheme which outperforms the array-level scheme. {  We also propose a channel decomposition method that enables effective ways for theoretically analyzing the ReRAM channel.} Simulation results show that the proposed SPI-aware adaptive detection and decoding schemes can approach the ideal performance with three quantization bits, with only one decoding iteration.

\end{abstract}


\section{Introduction}

The non-volatile memory (NVM) technologies have been developed rapidly which offer lower power consumption, faster access time, and better reliability than volatile memories. The resistive random-access memory (ReRAM) is widely considered the most promising NVM technology due to its superior features of high-density, high-speed, and long data retention time \cite{overview}. However, the crossbar array structure of ReRAM incurs a severe problem known as the sneak path, which will decrease the measured resistance value of the memory cell, and hence degrade the reliability of data recovery \cite{memristor}. As the sneak path interference (SPI) is data-dependent and also correlated within a memory array, it is a difficult problem to be tackled for the ReRAM crossbar array.

Several literature works have been proposed to combat the SPI from channel coding and detection perspectives. Sophisticated SPI models are considered in \cite{detection}, based on which the data stored in each memory cell is detected separately by treating the SPI as data-independent and uncorrelated noise. Data detection schemes for the SPI by using the side information provided by the pilot cells are proposed by \cite{pilot}. Across-array coding schemes are further developed by \cite{performance, across}, which distribute an error correction code (ECC) codeword to multiple memory arrays to reduce the channel correlation caused by the SPI. However, all these works are derived for the ReRAM channel without quantization.

Channel quantization is critical for supporting effective error correction coding for the data storage and communication systems. The one-bit quantizer is essentially the threshold detector ({\it i.e.} hard-output channel detector), while the multi-bit quantizer can be considered as the soft-output channel detector since it enables the generation of the log-likelihood ratio (LLR) values of the channel coded bits for decoding.  For the NVM channels, different criteria have been proposed to design the channel quantizers, based on the channel capacity, the channel cutoff rate, the Polyanskiy-Poor-Verdu finite-length performance bound, and so on \cite{on}. Among them the maximizing-mutual-information (MMI) criterion is the most widely used criterion for designing the channel quantizer for various NVM channels \cite{soft, on}.

In this work, we first propose a quantized channel model of ReRAM, based on which we design both the one-bit and multi-bit channel quantizers by maximizing the MI of the channel. A key channel parameter that affects the quantizer design is the sneak path occurrence probability (SPOP) of the memory cell. We first use the average SPOP calculated statistically to design the quantizer, which leads to the same channel detector over different memory arrays. {  To further improve the channel detection performance, we adopt the SPOP estimated separately for each memory array for the quantizer design, which is generated by an effective channel estimator and through an iterative detection and decoding (IDD) scheme for the low-density parity-check (LDPC) coded ReRAM channel.} This results in an array-level SPI-aware adaptive detection and decoding approach. Moreover, since the previous work \cite{across,pilot} observed that there is a strong correlation of the SPI that affects memory cells in the same rows/columns than that affecting cells in different rows/columns, we further derive a column-level scheme which outperforms the array-level scheme. {  We also propose a channel decomposition
method that enables effective ways for theoretically analyzing the ReRAM channel.} Simulation results show that our SPI-aware adaptive detection and decoding schemes can approach the ideal performance with three quantization bits, with only one decoding iteration. We remark that our proposed schemes can be used together with the prior art coding schemes for ReRAM \cite{performance, across} to achieve better performance.

\section{Quantized Channel Model of ReRAM}

In this work, we consider an $M \times N$ memory array where the ReRAM cell that locates at the intersection of row $i$ and column $j$ is denoted as cell $(i,j)$, with $i\in\{1,..., M\}$ and $j\in\{1,..., N\}$. Each memory array can store an $M \times N$ binary data matrix $X = [x_{i,j}]_{M \times N}$, with $x_{i,j} \in \{0,1\}$ being stored in cell $(i,j)$.  During writing, if $x_{i,j}=0$, the cell is programmed to a high resistance state
(HRS) with the nominal resistance value of $R_0$; if $x_{i,j}=1$, the cell is turned to a low resistance state (LRS) with the nominal resistance value of $R_1$. During reading, a certain voltage is applied to the target cell to measure its resistance. The memory reading may become unreliable due to the existence of the SPI. A sneak path is an undesired path that originates from and returns to the measured cell, and a direct effect of the SPI is a reduction of the measured resistance value. Hence, it is harmful only when the target cell is in HRS. The most popular method to mitigate the SPI is to introduce a cell selector to each memory cell, which is an electrical device that allows current to flow only in one direction \cite{memristor}. However, cell selectors are also prone to failures, leading to the re-occurrence of the SPI. The failure of the selectors can be assumed to be independent and identically distributed ({\it i.i.d.}) with probability $p_f$ \cite{detection, pilot}. Following the literature \cite{detection, pilot}, we take $p_f=0.001$ in this work. In addition, the random variation of the resistances of ReRAM may also result in the read decision errors of ReRAM cells \cite{overview, across}.

\begin{figure}[t]
	\centering
	\includegraphics[width=8cm]{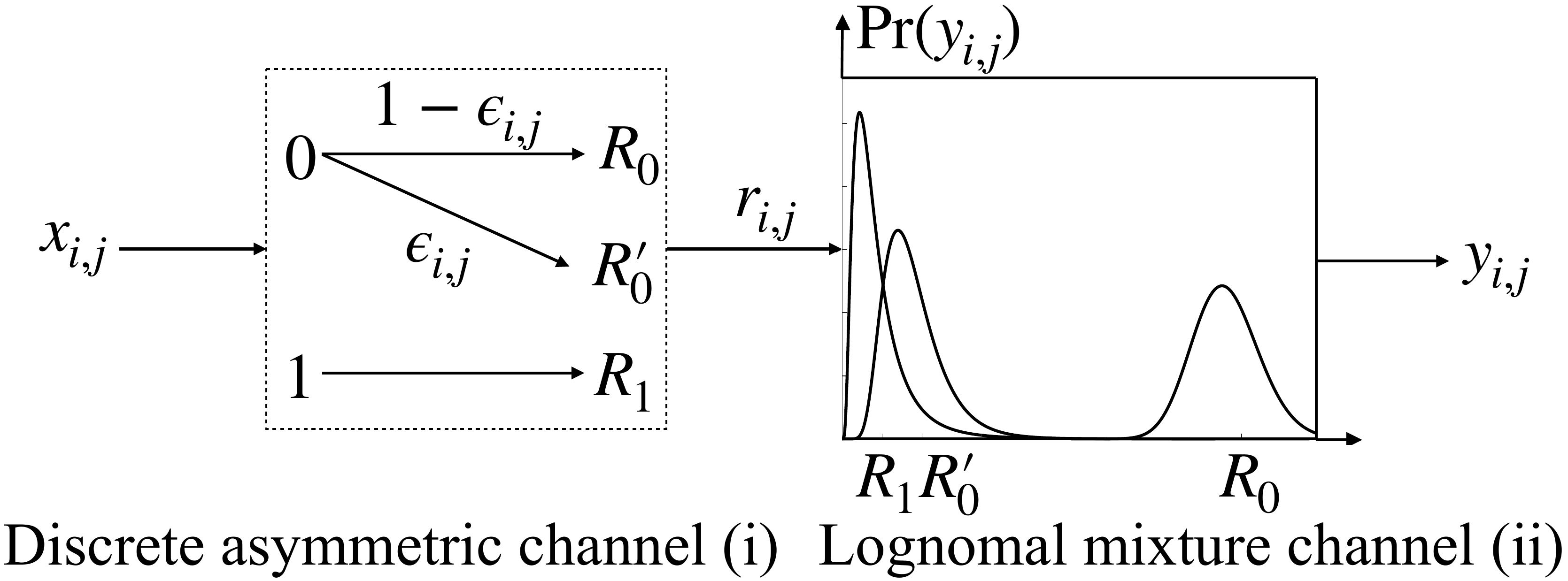}
	\caption{\ Cascaded channel model of resistive memory arrays.}
	\label{channel1}
\end{figure}

The SPI and random resistance variations of ReRAM are incorporated comprehensively by a cascaded channel model proposed by \cite{across}. In this model, as illustrated by Fig. \ref{channel1}, a discrete asymmetric channel is used to model the SPI of the crossbar array. Here, $\epsilon_{i,j}$ is the SPOP of cell $(i,j)$, {\it i.e.}, $\epsilon_{i,j}=\Pr(e_{i,j}=1|x_{i,j}=0)$, with a boolean variable $e_{i,j}$ being the SPI indicator. That is, $e_{i,j}=1$ if the cell $(i,j)$ is affected by SPI, and $e_{i,j}=0$ otherwise. The resistance value incorporating the SPI is given by $r_{i,j}=R_1$ for $x_{i,j}=1$, and $r_{i,j}=\frac{1}{\frac{1}{R_0}+\frac{e_{i,j}}{R_p}}$ for $x_{i,j}=0$, where $R_p$ is the parallel resistance introduced by the SPI. We thus have $r_{i,j} \in\{R_0, R_0', R_1\}$, with $R_0'=\frac{1}{\frac{1}{R_0}+\frac{1}{R_p}}$. Follow the literature \cite{detection,pilot,performance,across}, we take $R_0=1000\Omega$, $R_1=100\Omega$, and $R_p=250\Omega $, and hence $R_0'=200\Omega$.

Following the discrete asymmetric channel, a lognormal mixture model is further used to model the random resistance variations of ReRAM cells. The resistance $r_{i,j}$ is assumed to be deviated from its nominal values $R_0, R_0', R_1$ with variances $\sigma_{0}^2, \sigma_{0'}^{2}, \sigma_{1}^2$. Following literature \cite{cascaded}, it is further assumed that $\frac{\sigma_{0}}{R_0}=\frac{\sigma_{0}'}{R_0'}=\frac{\sigma_{1}}{R_1}=\frac{\sigma}{\mu}$. The resulting signal $y_{i,j}$ follows the lognormal distribution, {\it i.e.}, $\ln (y_{i,j})\sim \mathcal{N} (\mu_{i,j}, \sigma_{i,j}^2)$, where the values of $\mu_{i,j}, \sigma_{i,j}$ can be expressed as functions of $R_0, R_0', R_1$ and $\frac{\sigma}{\mu}$ \cite{across}. In the simulations, we vary the value of $\frac{\sigma}{\mu}$ to model the influence of different device designs and process variations on the resistance distributions of ReRAM cells.

\begin{figure}[t]
	\centering
	\includegraphics[width=6cm]{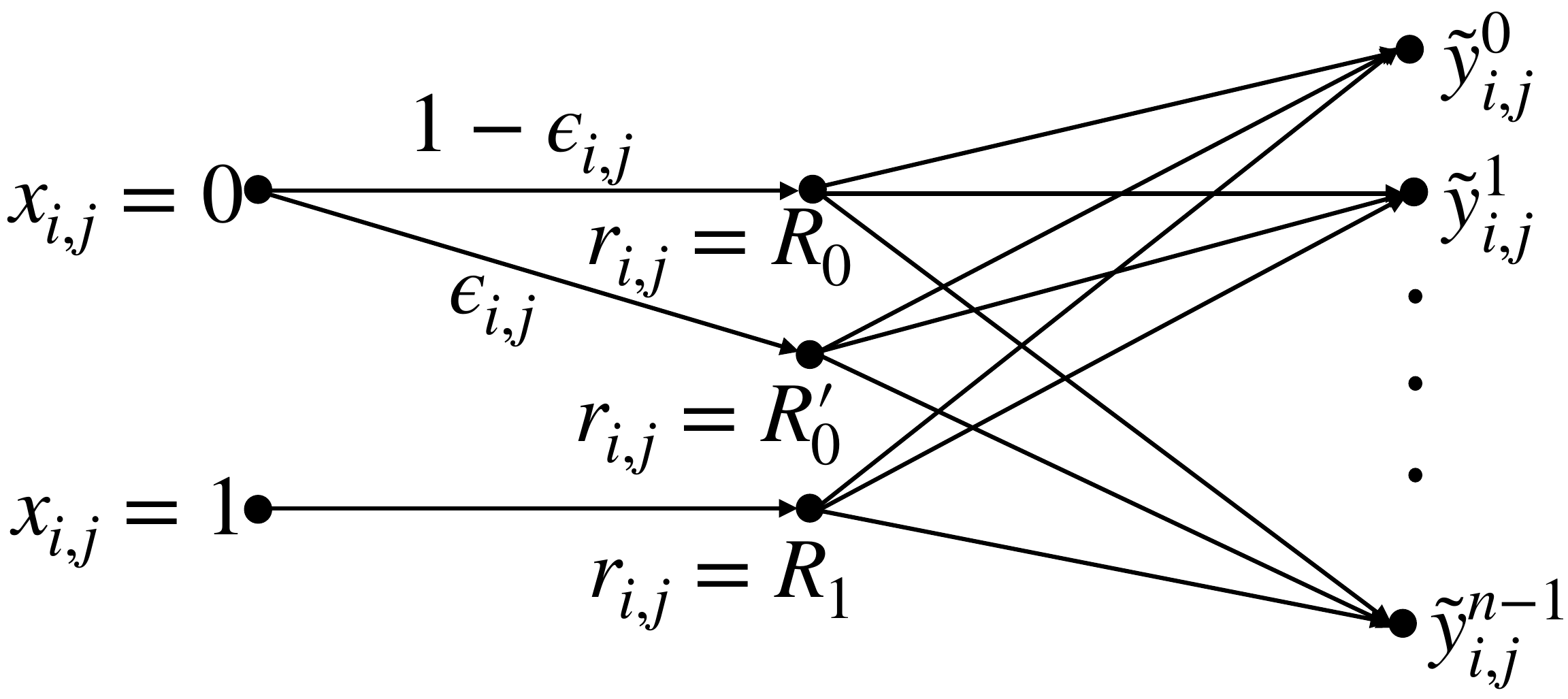}
	\caption{Quantized channel model of resistive memory arrays.}
	\label{channel2}
\end{figure}

Based on the above described cascaded channel model, we derive a quantized channel model of ReRAM. It serves as the basis for the design of channel quantizers. As shown by Fig. \ref{channel2}, a $p$-bit quantizer maps the channel output signal $y_{i,j}$ into $n = 2^p$ quantized outputs $\tilde{y}_{i,j}$, with $\boldsymbol{w}=[w_1,...,w_{n-1}]$ being the quantization boundaries. For simplicity, we use $x,y,r,\epsilon$ instead of $x_{i,j}, y_{i,j}, r_{i,j}, \epsilon_{i,j}$ in our subsequent derivations. The transition probabilities of the quantized ReRAM channel are then given by
\begin{equation}
\begin{cases}
P(\tilde{y}^k|x=0)=(1-\epsilon)P(\tilde{y}^k|r=R_0)+\epsilon P(\tilde{y}^k|r=R'_0)\\
P(\tilde{y}^k|x=1)=P(\tilde{y}^k|r=R_1)
\end{cases}.
\end{equation}
Given the cumulative distribution function (CDF) of the lognormal distribution $F(x,\mu,\sigma)=\frac{1}{2}[1+{\rm erf}(\frac{{\rm ln} x-\mu}{\sigma \sqrt{2}} )]$,
with ${\rm erf}(z)=\frac{2}{\pi} \int_0^z {\rm e}^{-t^2} dt $, and for $r=R_1$, we have
\begin{equation}
\begin{cases}
P(\tilde{y}^0|r=R_1)=F(w_1,\mu_1,\sigma_1)\\
P(\tilde{y}^k|r=R_1)=F(w_{k},\mu_1,\sigma_1)-F(w_{k+1},\mu_1,\sigma_1), 0<k<n-1\\
P(\tilde{y}^{n-1}|r=R_1)=1-F(w_{n-1},\mu_1,\sigma_1)
\end{cases}.
\end{equation}
The quantized channel transition probabilities for the cases of $r=R'_0$ and $r=R_0$ can be derived similarly.

\section{Maximizing Mutual Information (MMI)-Based Quantization For ReRAM Channels}

\subsection{One-Bit and Multi-Bit Quantization}

Consider a $p$-bit quantization, we use $w^*$ to denote the optimized boundary for one-bit quantization with $p=1$, and use $\boldsymbol{w^*}=[w^*_1,...,w^*_{n-1}]$ to denote the optimized multi-bit quantization boundaries for $p>1$. Consider the input data distribution ${\rm Pr}(x = 1) = q, {\rm Pr}(x = 0) = 1-q$, the MI of the proposed quantized ReRAM channel is given by
\begin{align}
\nonumber &I(X;Y)=\sum_{k=0}^{n-1} [ qP(\tilde{y}^k|x=1) {\rm log_2} P(\tilde{y}^k|x=1)\\
& + (1-q)P(\tilde{y}^k|x=0) {\rm log_2} P(\tilde{y}^k|x=0) ] -\sum_{k=0}^{n-1}P(\tilde{y}^k) {\rm log_2} P(\tilde{y}^k), \nonumber
\end{align}
where $ P(\tilde{y}^k)=(1-q) P(\tilde{y}^k|x=0)+q P(\tilde{y}^k|x=1)$.

For one-bit quantization, it is easy to show that $I(X;Y)$ has only one global maximum point. We can then obtain $w^*$ by solving $\frac{dI(X;Y)}{dw} = \frac{dH(Y)}{dw}-\frac{dH(Y|X)}{dw} = 0$. We thus have
We thus have $\frac{dH(Y)}{dw}=-\frac{dA}{dw} {\rm log_2} \frac{A}{1-A}$ and $\frac{dH(Y|X)}{dw}=-\sum_{k=0}^{n-1}  \left(qB+(1-q)C \right)$,
with $A=(1-q)\left[ \epsilon F(w,\mu'_0,\sigma'_0)+ (1-\epsilon) F(w,\mu_0,\sigma_0) \right]+ q F(w,\mu_1,\sigma_1)$, $B=\frac{1}{\rm ln 2} \frac{dP}{dw} (\tilde{y}^k|x=1) + \frac{dP}{dw}(\tilde{y}^k|x=1) {\rm log_2} P(\tilde{y}^k|x=1)$, and $C=\frac{1}{\rm ln 2} \frac{dP}{dw}(\tilde{y}^k|x=0) + \frac{dP}{dw}(\tilde{y}^k|x=0) {\rm log_2} P(\tilde{y}^k|x=0)$.

For multi-bit quantization, it is not possible to use analytical method to find the optimum quantization boundaries $\boldsymbol{w^*}$. Following \cite{dynamic}, we use dynamic programing (DP) to find the optimum $\boldsymbol{w^*}$. In particular, we first obtain $m$ boundaries: $s_1,..., s_{m}$ by uniformly dividing the interval $[s_1,s_{m}]$, with $m\gg n$, where $s_1$ and $s_{m}$ are predetermined to preserve most of the MI. Denote the problem of finding $\{w_1,...,w_{n-1}\}$ from $\{s_1,...,s_{m}\}$ as $P(m,n)$, with cost function $\Psi(m,n)=H(X|Y)$. Let $\Psi^*(m,n)$ be the cost function of the optimal solution, we have $\Psi^*(m,n)=\mathop{\min}_{n-1\le \lambda_{n-1}\le m} \Psi^*(\lambda_{n-1},n-1)+\psi(s_{\lambda_{n-1}+1},s_m)$,
where $\{\lambda^*_1,..., \lambda^*_{n-1}\}$ collects the indices of $\{s_1,..., s_m\}$ such that $\{s_{\lambda_1^*},..., s_{\lambda^*_{n-1}}\}$ is the optimal solution of $P(m,n)$, and $\psi(w_{a},w_{b})=\sum_{k=a}^{k=b}-\frac{1}{2}\sum_{g=0}^{1} P(y_k|x=g)\log \frac{P(y_k|x=g)}{2P(y_k)}$. The optimal solution of $P(m,n)$ can be obtained by solving $P(\lambda_{n-1},n-1)$ in a recursive way\cite{on}.

\subsection{Log-Likelihood Ratio (LLR) Calculation}\label{sec:llr}

Given the optimized quantization boundaries $\boldsymbol{w^*}=[w_1^*, w_2^*,...,w_{n-1}^*]$ derived above, the log-likelihood ratio (LLR) value of $x_{i,j}$ can be calculated as follows. For $w_k^*\textless y_{i,j}\textless w_{k+1}^*, k=1,...,n-2$,
\begin{align}
\nonumber &L(x_{i,j}|y_{i,j},\boldsymbol{w^*})=\log \frac{P(\tilde{y}^{k}|x_{i,j}=0)}{P(\tilde{y}^{k}|x_{i,j}=1)}+ \log \frac{P(x_{i,j}=0)}{P(x_{i,j}=1)}\\
&=\log \frac{D+E}{F(w_k^*,\mu_1,\sigma_1)-F(w_{k+1}^*,\mu_1,\sigma_1)}+ \log \left(\frac{1-q}{q}\right), \label{eq:rll}
\end{align}
with $D=\epsilon \left[ F(w_k^*,\mu'_0,\sigma'_0)-F(w^*_{k+1},\mu'_0,\sigma'_0)\right]$ and $E=(1-\epsilon) [F(w_k^*,\mu_0,\sigma_0)-F(w^*_{k+1},\mu_0,\sigma_0)]$.
The LLRs for the cases of $y_{i,j}\textless w_1^*$ and $y_{i,j}\textgreater w_{n-1}^*$ can be calculated in a similar way.

\section{SPI-Aware Adaptive Detection and Decoding}

\subsection{Array-Level Adaptive Detection and Decoding Scheme}

\begin{figure}[t]
	\centering
	\includegraphics[width=6cm]{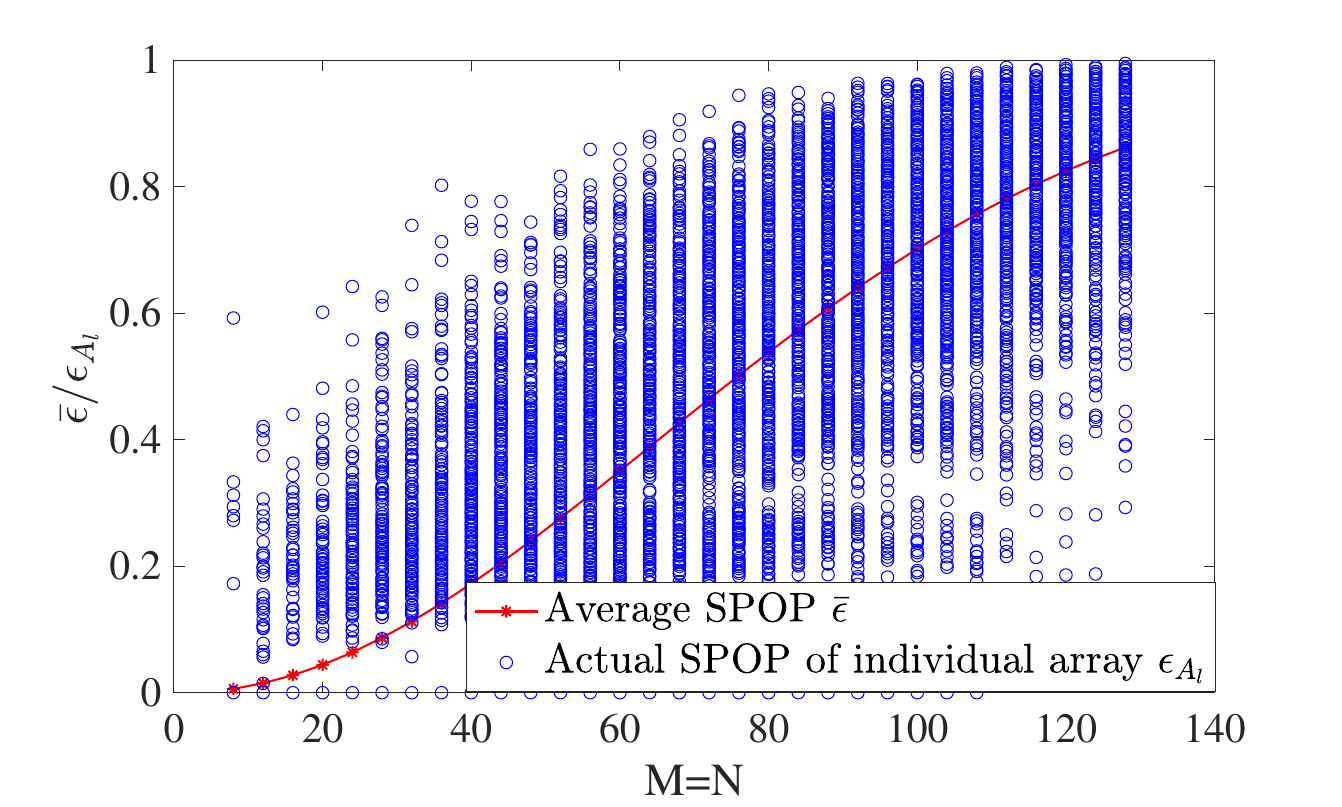}
	\caption{Average SPOP versus actual SPOP of ReRAM arrays.}
	\label{system}
\end{figure}

The SPOP $\epsilon$ is a key parameter that affects the quantizer design. Since it is an unknown channel parameter before channel detection and decoding, we can approximate it using the average SPOP $\bar{\epsilon}$ over different arrays, which can be calculated statistically as  $\bar{\epsilon}=1-\sum_{u=0}^{M-1} \sum_{v=0}^{N-1} \binom{M-1}{u} \binom{N-1}{v} q^{u+v} (1-q)^{M-1-u+N-1-v} (1-p_f q)^{uv} $\cite{performance}. This leads to the same channel detector over different memory arrays. However, our further investigation shows that the actual SPOPs of different arrays diverge far away from the average SPOP $\bar{\epsilon}$ (see Fig. 3). {  It is due to the data-dependent property of the SPI.}

Inspired by the above analysis, we adopt the SPOP estimated separately for each memory array for the quantizer design. We use $\epsilon_{A_l}$ to denote the SPOP of the $l$-th array. {  It is generated by an effective channel estimator and through an IDD scheme for the LDPC coded ReRAM channel, which is illustrated by Fig. \ref{system}.} In the fiture, the LDPC encoder encodes a binary user data word $\boldsymbol{x}$ into a codeword $\boldsymbol{c}$ which is then stored in the memory array. During the first round of detection, we adopt the average SPOP $\bar{\epsilon}$ to design a $p$-bit quantizer to quantize the resistance $\boldsymbol{y}$ measured from the memory cell. The LLR calculator produces the channel LLRs according to (\ref{eq:rll}), which are then sent to the LDPC decoder. After LDPC decoding, the channel estimator estimates the SPOP for each array separately using the LDPC decoded bits. The newly estimated SPOP $\hat{\epsilon}_{A_l}$ is then sent back to the quantizer to generate new quantization boundaries $\boldsymbol{w^*}$ for the $l$-th array, for the next round of detection and decoding. This results in an array-level adaptive detection and decoding scheme.

{  In this work, we propose an effective channel estimator based on the following key features of the ReRAM channel: only the detection of HRS cells is affected by the SPI. Moreover, the resistance of the HRS cell will be reduced significantly once being affected by the SPI. This can be clearly seen from the resistance distributions of the SPI-affected HRS cell (with mean value $R_0'$) and the SPI-free HRS cell (with mean value $R_0$) illustrated by Fig. 1(ii).}
Hence our channel estimator estimates $\hat{\epsilon}_{A_l}$ as follows. Let $N_{R0'}$ and $N_{R0}$ denote the number of SPI-affected and SPI-free HRS cells in a memory array, respectively. After LDPC decoding, we find out all the memory cells that are decoded to be in the HRS ({\it i.e.} decoded to be `0's). We then go back and check the measured resistance value $\boldsymbol{y}$ of the HRS cell. If it is less than a predetermined resistance threshold $R_{th}$, the cell is considered to be an SPI-affected cell. Otherwise, it is considered to be SPI-free. {  The $R_{th}$ is determined by minimizing the overlapping region of the distributions of the $R_0$ and $R_0'$ cells: $R_{th}=\mathop{\min}_{R_0' \le R \le R_0} [1-F(R,\mu_0',\sigma_0')+F(R,\mu_0,\sigma_0)] $. The SPOP $\hat{\epsilon}_{A_l}$ can then be estimated as $\hat{\epsilon}_{A_l}=N_{R0'}/(N_{R0'}+N_{R0})$.}

\begin{figure}[t]
	\centering
	\includegraphics[width=6cm]{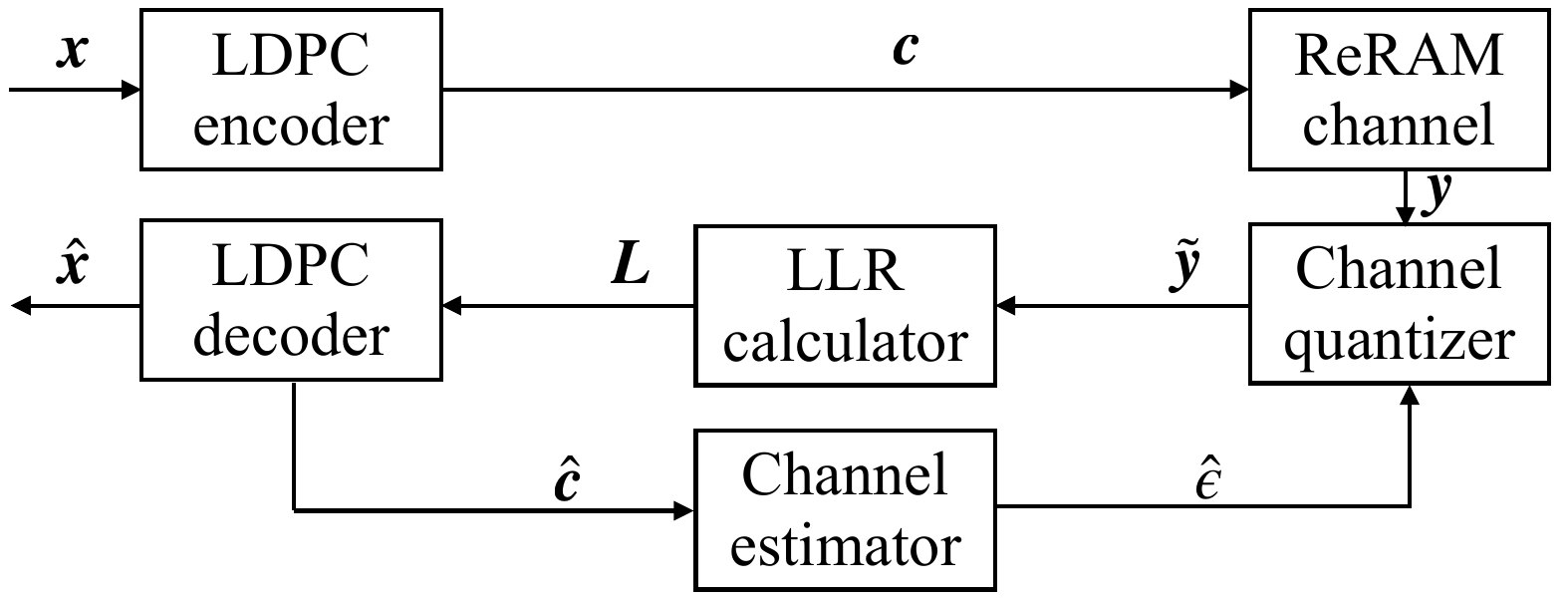}
	\caption{\ LDPC coded ReRAM system with adaptive and iterative detection and decoding {  and channel estimation.}}
	\label{system}
\end{figure}

\subsection{Column-Level Adaptive Detection and Decoding Scheme}

The above described adaptive and IDD scheme is at the array level. Note that in the typical memory circuit design, cells on the same wordline/column share the same sensing amplifier ({\it i.e.} threshold detector). Hence it is reasonable to adopt one detector for each column of the memory cells \cite{write}. More importantly, previous work [4], [6] observed that there is a strong correlation of the SPI that affects cells in the same rows/columns than that affecting cells in different rows/columns, and hence using the SPOP estimated separately for each column in an array for the quantizer design can achieve further performance gain. Here, we use $\epsilon_{C_j}$, $j=1,\cdots,N$, to denote the SPOP of the $j$-th column of an array. After the first round of decoding, the channel estimator estimates the SPOP separately for each column by using the LDPC decoded bits. The newly estimated SPOP $\hat{\epsilon}_{C_j}$ is then sent back to the quantizer to generate new quantization boundaries $\boldsymbol{w^*}$ for the $j$-th column of the array. This results in a column-level SPI-aware adaptive detection and decoding scheme. As will be shown in Section \ref{simulation}, it can achieve even better error rate performance than the array-level scheme.

{

During the implementation of the above proposed array-level and column-level schemes, the most time-consuming part is to use the estimated SPOP to re-design the quantizer and generate the new quantization boundaries $\boldsymbol{w^*}$ for each array/column. Accordingly, we propose to design the quantization boundaries $\boldsymbol{w^*}$ based on a given SPOP value offline, and store them in a lookup table. Through table lookup, we can avoid the online computational complexity of designing the quantizer for a newly estimated SPOP. Moreover, as will be shown in Section VI, taking one decoding iteration can already significantly improve the error rate performance, and there is no need to do more iterations. Therefore, the increase in latency of the proposed schemes is also very limited.
}

{
\section{Performance Analysis Based on Channel Decomposition}

Due to the data-dependency of the SPI, the SPOP (channel condition) varies from array to array. This creates significant challenges for the performance analysis and code design for the ReRAM channel, since conventional theoretical tools such as the density evolution (DE) and extrinsic information transfer (EXIT) chart analysis assume the channel to be stationary and data-independent. To address this issue, in this section,  we propose a channel decomposition method that decomposes the ReRAM channel into multiple {\it i.i.d.} channels (\textit{i.e.} data-independent and un-correlated channels), and demonstrate that the ReRAM channel performs almost the same as the mixture of the decomposed channels. 

In particular, if the SPOP of each array is known, the decoder considers the channels associated with different arrays as {\it i.i.d.} channels, where all the memory cells within each array are corrupted by the same SPOP $\epsilon$. Meanwhile, the SPOP $\epsilon$ varies from array to array. This is incorporated into the LLR calculation described by Section~\ref{sec:llr}. We denote such an {\it i.i.d.} channel as $W_{\epsilon}$. Therefore, the ReRAM channel associated with multiple memory arrays with varying $\epsilon, 0\leq \epsilon\leq 1$, can be considered as a mixed channel describe by
 $\sum_{\epsilon} Q_{\epsilon} W_{\epsilon}$, where $Q_{\epsilon}$ is the probability of an array with SPOP $\epsilon$. The decoding bit error rate (BER) of ReRAM can then be estimated by
 \begin{align}
    P^{ReRAM}_e \approx\sum_{\epsilon} Q_{\epsilon}f(\epsilon)\label{eq:BER}
 \end{align}
 where $f(\epsilon)$ is the decoding BER of $W_{\epsilon}$.

As the SPOP is determined by the number of selector failures (SFs) associated with the LRS cells, such an SF is named as an active SF. It can be proven that most of the SPI-affected cells are HRS cells that do not fall in the same rows or columns with the active SFs, if the number of active SFs within array, denoted by $N_{SF}$, is small. The probability of such an HRS cell being affected by the SPI is $1-(1-q^2)^{N_{SF}}$\cite{performance}. Then the law of large numbers guarantees that the SPOP of a given array approaches $1-(1-q^2)^{N_{SF}}$ if the array size is sufficiently large. Moreover, it can be easily shown that the probability of an array with $N_{SF}$ active SFs is given by
\begin{equation}
	q_{N_{SF}}=\binom{M}{N_{SF}} \binom{N}{N_{SF}} N_{SF} ! (1-p_f q)^{MN-N_{SF}} (p_f q)^{N_{SF}}.\nonumber
\end{equation}
Since in this case $Q_{\epsilon}=q_{N_{SF}}$, (\ref{eq:BER}) can be rewritten as
\begin{align}
	\nonumber P^{ReRAM}_e  \approx& \sum_{N_{SF}} \binom{M}{N_{SF}} \binom{N}{N_{SF}}N_{SF}!(1-p_f q)^{MN-N_{SF}}\\
	\label{eq:BER2} &\times (p_f q)^{N_{SF}}f\left(1-(1-q^2)^{N_{SF}}\right).
\end{align}
As $q_{N_{SF}}$ decreases quickly with the increase of $N_{SF}$, we only need to consider the first few terms when calculating (\ref{eq:BER2}).

\begin{figure}[b]
	\centering
	\includegraphics[width=7cm]{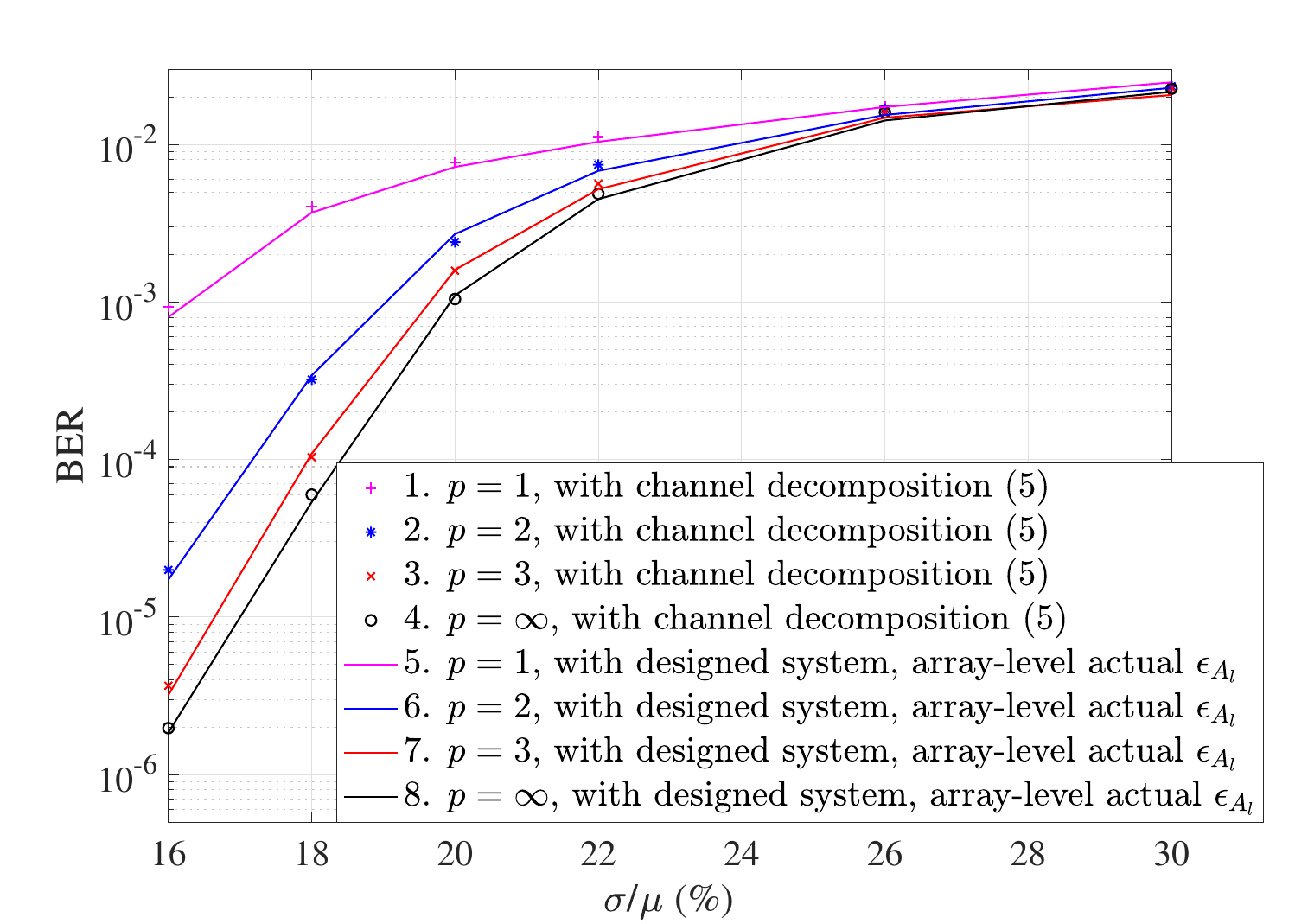}
	\caption{{  BER comparison of LDPC coded ReRAM system designed based on the array-level actual SPOP ${\epsilon}_{A_l}$ and those obtained by channel decomposition, with $M\times N=32\times 32$.}}
	
\end{figure}

For LDPC coded ReRAM system with array size $32\times 32$ (details of the LDPC code are given in Section \ref{simulation}), Fig. 5 compared the simulated decoding BERs for the system designed based on the array-level actual SPOP ${\epsilon}_{A_l}$ and those obtained by channel decomposition (\ref{eq:BER2}), where $N_{SF}\leq 3$ is considered and $f(\cdot)$ is obtained by simulation.  Observe that the BERs for the two cases match well with each other, for both with and without quantization. Note that $f(\cdot)$ can also be obtained using theoretical tools such as DE and EXIT chart with infinite code length assumption. Therefore, the above proposed channel decomposition clears the obstacles and enables effective ways for theoretically analyzing the ReRAM channel. Following the idea of channel decomposition, in this work, we design and optimize the channel quantizer/detector through the channel estimator and IDD. We leave the further performance analysis and code design for the ReRAM channel as our future work.
}

\section{Simulation Results} \label{simulation}

In the simulations, we adopt a systematic LDPC code generated by the progressive-edge-growth (PEG) algorithm, with code length $1024$ and code rate $R=0.88$. Note that other ECCs can also be used for the proposed schemes.

\begin{figure}[t]
	\centering
	\includegraphics[width=7.1cm]{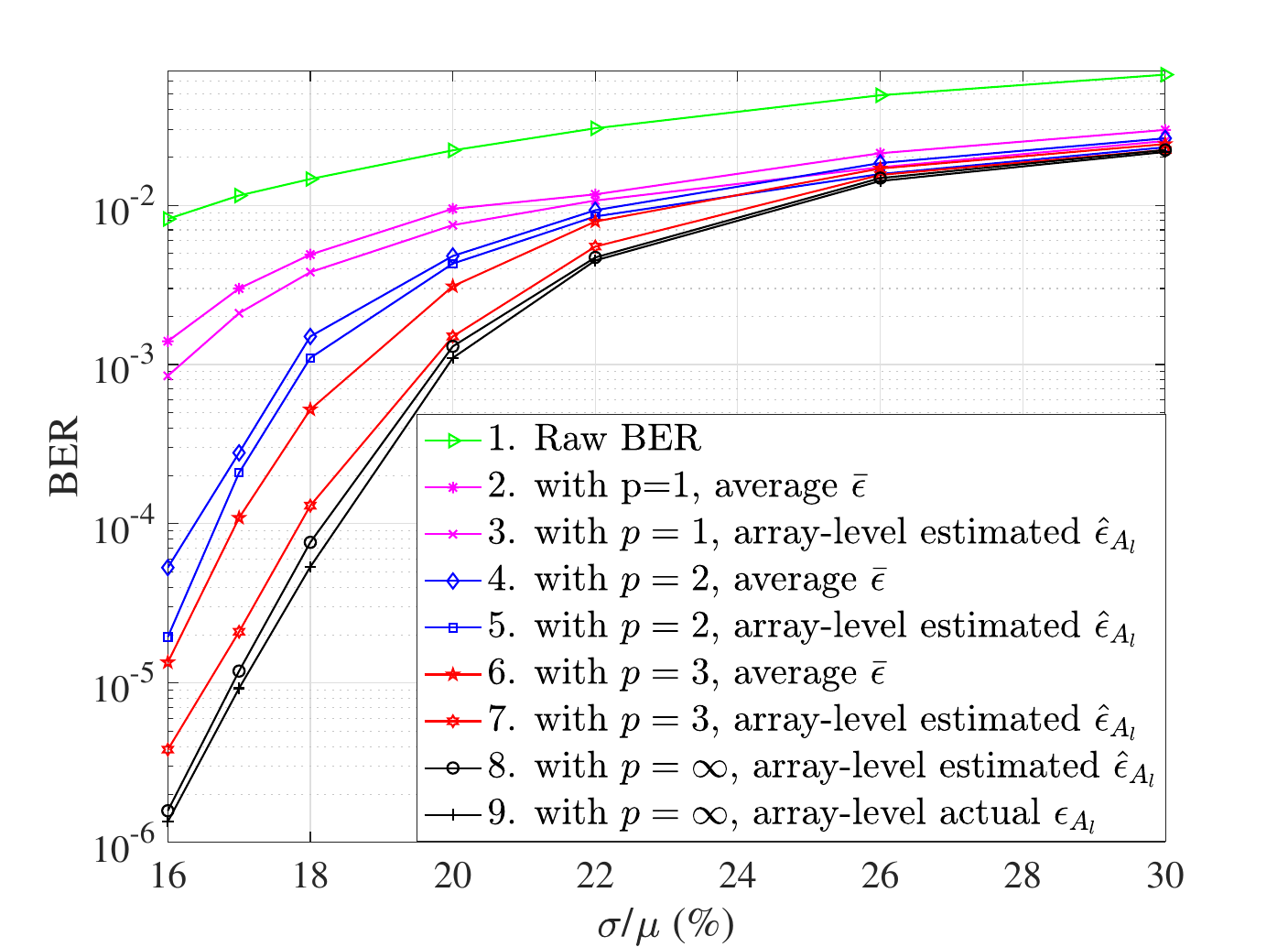}
	\caption{\ BER comparison of different array-level schemes with schemes using average SPOP $\bar{\epsilon}$, with $M\times N=32\times 32$.}
	\label{ber1}
\end{figure}

Fig. \ref{ber1} shows the BER performance of the LDPC coded system with different quantization/detection schemes, with the array size of $M \times N=32\times 32$. In the figure, {  Curve 1 shows the channel raw BERs obtained by making hard decisions based on the threshold $R_{th}^0=\mathop{\min}_{R_1 \le R \le R_0'} [1-F(R,\mu_1,\sigma_1)+F(R,\mu_0',\sigma_0')]$.} Curves 2, 4, and 6 illustrate the BERs with the quantizers being designed using the average SPOP $\bar{\epsilon}$ without decoding iteration, for $p=1$, $p=2$, and $p=3$, respectively. The performance of the proposed array-level adaptive detection and decoding scheme,  with different quantization bits, is shown by Curves 3, 5, and 7, respectively. Only one decoding iteration is carried out for these curves. Curve 8 illustrates the BERs of the system without quantization ({\it i.e.} $p=\infty$), and the LLRs are calculated using the array-level estimated SPOP $\hat{\epsilon}_{A_l}$. Curve 9 shows the BERs of the unquantized system with the LLRs being generated based on the array-level actual SPOP ${\epsilon}_{A_l}$. Hence it is an ideal case that serves as a reference. Observe from the figure that the BERs improve with the increase of the number of quantization bits $p$, for both cases without and with the proposed array-level scheme. With only one decoding iteration, our proposed array-level scheme achieves significant BER gain over the scheme using the average SPOP $\bar{\epsilon}$, for different numbers of quantization bits. With three-bit quantization, our proposed scheme approaches the performance of the ideal case where the LLRs are generated using the actual SPOP ${\epsilon}_{A_l}$ without quantization. The negligible gap between Curves 8 and 9 indicates that no significant performance gain will be obtained with more decoding iterations.

\begin{figure}[t]
	\centering
	\includegraphics[width=7.1cm]{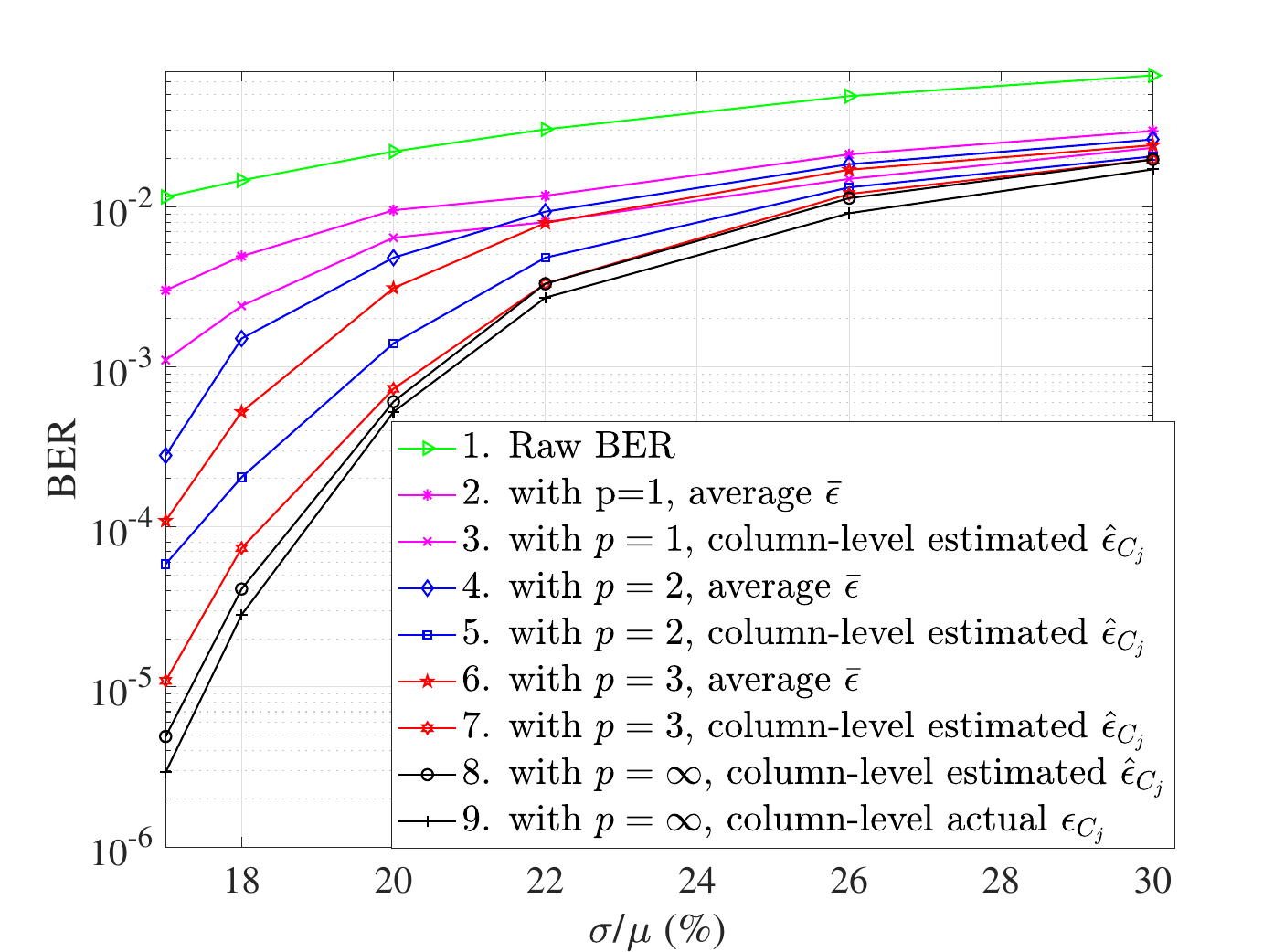}
	\caption{\ BER comparison of different column-level schemes with schemes using average SPOP $\bar{\epsilon}$, with $M\times N=32\times 32$.}
	\label{ber2}
\end{figure}

Similar BER evaluations are performed for our proposed column-level scheme, and the results are shown in Fig. \ref{ber2}. Observe that the relative performance of different curves remains the same. Moreover, the column-level scheme achieves obvious performance gain over the array-level scheme. We also evaluate the performance of the proposed schemes over larger memory arrays, and similar observations are obtained.

\vspace{-4mm}

\section{Conclusion}

We have proposed novel SPI-aware adaptive detection and decoding schemes to tackle the SPI of the ReRAM crossbar array. We have first developed a quantized channel model of ReRAM, based on which we designed both the one-bit and multi-bit channel quantizers by maximizing the MI of the channel. {  We have then proposed the array-level and column-level adaptive and iterative detection and decoding scheme, through which a channel estimator can estimate the SPOP at the array/column level accurately, and the latter is used to update the design of the channel quantizer for the next round of detection and decoding. We have also proposed a channel decomposition
method that enables effective ways for theoretically analyzing the ReRAM channel. We leave the further performance analysis and code design for the ReRAM channel as our future work.}

\end{document}